\documentclass[aps,prd,showpacs,preprint,amsfonts]{revtex4-1}
\usepackage{epsfig}
\usepackage{graphics}
\usepackage{amsmath}

\newcommand{\Eq}[1]{Eq.~(\ref{#1})}
\newcommand{\Eqs}[1]{Eqs.~(\ref{#1})}

\newcommand{\be}{\begin{equation}}
\newcommand{\ee}{\end{equation}}
\newcommand{\bea}{\begin{eqnarray}}
\newcommand{\eea}{\end{eqnarray}}
\newcommand{\half}{\frac{\scriptstyle 1}{\scriptstyle 2}}
\newcommand{\fourth}{\frac{\scriptstyle 1}{\scriptstyle 4}}

\begin{document}

\author{Jacob D. Bekenstein}
\affiliation{Racah Institute of Physics, Hebrew University of
Jerusalem, Jerusalem 91904, Israel\\}\date{\today}
\title{If vacuum energy can be negative, why is mass always positive?: Uses of the subdominant trace energy condition}
\begin{abstract}
Diverse calculations have shown that a relativistic field confined to a cavity by well defined boundary conditions can have a negative Casimir or vacuum energy.  Why then can one not make a finite system with negative mass by confining the field in a some way?  We recall, and justify in detail, the not so familiar subdominant trace energy condition for ordinary (baryon-electron nonrelativistic) matter.  With its help we show, in two ways, that the mass-energy of the cavity structure necessary to enforce the boundary conditions must exceed the magnitude of the negative vacuum energy, so that all systems of the type envisaged necessarily  have positive mass-energy.   

\end{abstract}
\maketitle

\section{Introduction}
\label{sec:intro}

Casimir~\cite{Casimir} first showed that the vacuum energy of the electromagnetic field in the presence of two parallel and perfectly conducting plates must be negative.  The existence of a nontrivial Casimir energy for every confined quantum field is by now well documented, theoretically and experimentally, not only for the situation envisaged by Casimir, but also for quantum fields of various types confined to cavities of diverse shapes~\cite{Boyer,Lukosz,Feinberg,Dowker,Deutsch,Ford,Wolfram,Mostep}.  In particular, other examples of negative Casimir energies are known~\cite{Lukosz,Dowker,Deutsch,Wolfram}.

We never observe negative masses.  So what prevents us from using negative Casimir energy to make a physical system of finite size with zero or negative mass?  In particular, why can one not engineer the wall of the cavity of so light material that its mass $M_\mathrm{w}$ will not exceed the absolute value the Casimir energy? 

An easy but unsatisfying answer is the following.  
One observes that in the calculated examples of a field in a cavity of typical dimension $R$, the  Casimir energy turns out to be $\alpha\hbar c/R$ with $|\alpha|$ in the range $10^{-4}-10^{-1}$.  The wall, being of size $R$, sets an upper bound on the size of its constituent particles.  Each such particle must be  larger than its own Compton length $\hbar/\mu c$.  Thus $M_\mathrm{w}$ should be at least a few times $\hbar/c R$.  Hence the cavity mass-energy plus the Casimir energy should come out positive.  The unsatisfying side of such an argument is that no known \emph{principle} restricts the value of $\alpha$.  The Casimir energy is the regularized value of the infinite sum $\half (E_1+E_2+\cdots),$ where $E_j$ is the energy of the $j$-th mode of the field in the cavity.  Typically $E_j\sim \hbar c/R$ multiplied by a number which rapidly becomes much larger than unity as $j$ grows.  Why regularization of this divergent sum should always yield an $\alpha$ tiny on scale unity is not at all clear.  A largish $\alpha$ would, of course, nullify the above simplistic argument.
 
Existence of a system with negative (inertial or gravitating) energy, quite apart from seeming incongruous, would lead to well known conceptual problems.  To mention a little known example, black hole physics together with the second law of thermodynamics yields a universal bound on the ratio of entropy to total mass-energy of an ordinary system~\cite{Bek81}.  If the mass-energy of an entropy bearing system were negative, one could, by adding bits of matter to it, bring its mass very close to, but above zero.  The entropy to energy ratio could thus be made to exceed the stated bound~\cite{Wolfram}.   Examples such as this strengthen the feeling that the mass-energy of an isolated physical system cannot be negative or even arbitrarily near zero from above.

Bearing in mind our original example of a system, one surmises that quite apart from the quantum argument adduced above, the requirement that the wall be sufficiently rigid to withstand the suction exerted by the field (in its negative energy vacuum state) necessarily makes it massive enough to overcompensate for the negative Casimir energy~\cite{Wolfram}.  In Secs.~\ref{sec:way1} and~\ref{sec:way2} we develop two different formal argument in support of this idea for cavities of \emph{arbitrary shape} confining massless fields of several kinds. We assume that the wall is a classical structure which respects the subdominant trace energy  condition.  This little known condition is motivated and supported, in Sec.~\ref{sec:sub}, by two lines of evidence employing kinetic theory and field theory, respectively.  The bottom line is that quantum fields belonging to a broad class, when confined to a classical cavity, cannot force the total mass-energy to be negative.

We work mostly in flat four-dimensional spacetime (but some of our arguments, like that in Sec.~\ref{sec:way1}, are extendable to higher dimensions).  In formal arguments we sometimes refer to curved spacetime.  We choose the metric signature $(-,+,+,+)$.  In the whole of Sec.~\ref{sec:sub} we shall use units with $c=1$.  Greek tensor indices are spacetime indices, Roman indices, if from mid-alphabet, refer to spatial coordinates; if from the early alphabet they refer to group indexes, and if capitalized they refer to tetrad indices (range 0 to 3).    Unless otherwise specified, pairs of identical indices of all kinds appearing in the same term are  to be summed over their whole range (Einstein convention).

\section{the subdominant trace energy condition}
\label{sec:sub}
\subsection{Definition}
\label{sec:Def}

In this paper the cavity wall is regarded as made of ordinary (baryon-electron nonrelativistic) matter.  We denote the energy-momentum tensor of such matter by $\Theta_{\alpha\beta}$. We shall here justify the applicability to ordinary matter of a little appreciated energy condition, the \emph{subdominant trace energy condition}, namely
\be
|\Theta_i{}^i|<\Theta_{tt}\,.
\label{cond}
\ee
(We consistently write the energy conditions as strict inequalities, thus excluding the cases when they are saturated.)
\Eq{cond} says that the energy density $\Theta_{tt}$ (assumed positive) is claimed to exceed the magnitude of the trace of the stress tensor (spatial-spatial part of $\Theta_\mu{}^\mu$).  In covariant notation the subdominant energy condition is
\be
|\Theta_{\alpha\beta}(g^{\alpha\beta}+u^\alpha\,u^\beta)|<\Theta_{\alpha\beta}\,u^\alpha\,u^\beta\,,
\ee 
where $u^\alpha$ is any future-pointing 4-vector $u^\alpha$ satisfying $u^\alpha\,u_\alpha=-1$ and representing the velocity of an observer.
 
A commonly invoked energy condition applicable to classical matter is the \emph{strong energy condition} or SEC~\cite{HawkingEllis}, written covariantly as 
\begin{equation}
(\Theta_{\alpha\beta}-\half g_{\alpha\beta}\, \Theta_\mu^\mu)\,u^\alpha u^\beta > 0\,.
\label{cond2}
\end{equation}
Among many other applications, the SEC has served as basis for  a succinct proof, within general relativity, of the positivity of mass of any static singularity-free classical system~\cite{Bek75a}.
Substituting $u^\alpha=\delta^\alpha_t$ and $g_{\alpha\beta}\,u^\alpha u^\beta=-1$ into \Eq{cond2}  gives
\begin{equation}
\Theta_{tt}+\Theta_i{}^i>0.
\label{TT}
\end{equation}
This recovers condition (\ref{cond}) whenever $\Theta_i{}^i<0$, assuming, as is reasonable for classical matter, that the energy density is positive.  

Not every conceivable physical medium obeys \Eq{cond2}.  For example, the celebrated cosmological  ``dark energy'' can be modeled as an isotropic medium in which \emph{each} $\Theta_i{}^i$ is approximately $-\Theta_{tt}$; thus \Eq{cond2} fails for it.     And a free massive scalar field can also violate the strong energy condition~\cite{Bek75b}.  Accordingly, in their assessment of the status of the various energy conditions, Barcelo and Visser~\cite{Barcelo} declare the SEC as ``dead''.  Of course,  any rule in physics will break down if pushed beyond its realm of validity.  And SEC has a realm of validity which certainly includes ordinary matter.  In any case, in Sec.~\ref{sec:two_proofs} we shall \emph{not} make use of \Eqs{cond2}-(\ref{TT}), but only of the subdominant trace condition for the case $\Theta_i^i>0$, which is supported by the arguments collected in the rest of this section.
 
 Another common energy condition applicable to classical matter is the \emph{dominant energy condition} or DEC~\cite{HawkingEllis}, which in covariant notation  stipulates that  
\be
  \Theta_{\alpha\beta}\,u^\beta\,u^\alpha>0 ;\qquad \Theta_{\alpha\beta}\,u^\beta\ \Theta^\alpha{}_\gamma\, u^\gamma <0.
  \label{cond1}
\ee
This states that \emph{any} physical observer not only sees positive energy density, but also a \emph{timelike} Pointing vector (energy flux 4-vector), i.e., the energy flows only within the lightcone.

Let us assume, in addition to stationarity, that in the local inertial rest frame of the matter $\Theta_t{}^j=0$, e.g., no rotation.   Take the observer velocity of the form $u^\alpha= u^t \delta_t{}^\alpha+u^k \delta_k{}^\alpha$ (no sum over $k$). Assume also that $u^t\gg 1$, which in view of the normalization  $u^\alpha\,u_\alpha=-1$ can be true only if $|u^k|\gg 1$ as well (observer is rapidly moving with respect to the matter).  Then conditions (\ref{cond1}) boil down to
\be
\Theta_{tt}>0; \qquad  \Theta_{tt}> |\Theta_i{}^k|\ {\rm for\ all}\ i\ {\rm and}\ k.
\label{domin}
\ee
In particular, if we align the spatial coordinates locally so that $\Theta_i{}^k$ becomes diagonal, the second of the above results means that $ \Theta_{tt}> |\Theta_i{}^i|$ (no sum over $i$).  
The DEC can also fail: Casimir's energy between the parallel plates is negative.  Barcelo and Visser thus regard the DEC as ``moribund''.  But we stress again that there is a regime of validity of DEC which includes ordinary matter.

The subdominant trace energy condition (\ref{cond}) that we shall exploit below is an extension of conditions~(\ref{TT}) and  (\ref{domin}) asserting that the energy density $\Theta_{tt}$ is positive, and  dominates the  \emph{trace} $\Theta_i{}^i$ also when the latter is positive.   Condition (\ref{cond}) implies the SEC as well as the positivity of energy or weak energy condition (WEC) and the null energy condition (NEC)~\cite{HawkingEllis} (M. Visser~\cite{Visser}).

\subsection{Argument from kinetic theory}
\label{kinetic}

What evidence have we for the validity of the subdominant trace condition?  Recall that for a gas of free nonrelativistic particles $\Theta_{tt}\gg |\Theta_k{}^j|$ as a a rule. According to kinetic theory this reflects the fact that the matter's constituents (atoms, nuclei, etc.) move slowly on scale $c$.  As the matter constituents become relativistic, the diagonal stresses $\Theta_k{}^k$ (no sum) will, again according to kinetic theory, each approach $\frac{1}{3} \Theta_{tt}$ from below, so that condition (\ref{cond}) remains satisfied and approaches saturation in the ultra relativistic limit.  Thus for the whole regime of motions in matter with negligible interactions we find that the subdominant trace condition is satisfied also when $\Theta_i{}^i>0$.   

Interactions, particularly when strong, could change the above story.  For example, it has been argued by Zel'dovich~\cite{Zeldovich} that in fluid matter in which the inter-particle interaction is mediated by a massive vector field, the individual stresses $\Theta_k{}^k$ (again no sum) at high density asymptomatically approach $\Theta_{tt}$.  This obviously violates condition (\ref{cond}) and so Zel'dovich and Novikov in their influential textbook~\cite{ZeldovichNovikov} do not consider this a  reliable energy condition. However, the said example is certainly not relevant to the question at hand, or indeed for generic ordinary matter which is a low density and low energy medium.   In addition the cavity wall---a solid---will generally have anisotropic $\Theta_i{}^j$ and  thus be very different from the hot isotropic matter envisaged by Zel'dovich.  However, one must go beyond these apologies and check the validity of condition (\ref{cond}) for ordinary matter under fairly general circumstances, and we turn now to the field theory arguments supporting it.

\subsection{Argument from field theory}

In this subsection the subdominant trace condition is supported by arguments based on a field theoretic description of ordinary matter.  Sec.~\ref{sec:basic} works out, at the classical level, the  relation between the trace of the energy momentum tensor and mass terms for theories possessing a liberalized form of conformal (or scale) symmetry.  The next three subsections show that all the ingredients of ordinary matter comply with this symmetry and thus comply with the above relation.   Sec.~\ref{sec:proof} takes up the Weyl (conformal) anomaly issue raised by field quantization, and goes on to infer the said energy condition.  The reader who finds the energy condition (\ref{cond}) palatable outright can skip directly to Sec.~\ref{sec:two_proofs}. 

We work directly with the energy momentum tensor as derived from the action for various fields of relevance.  The total action $S$ is a functional of a generic spacetime metric $g^{\mu\nu}$ (regarded as different from $\eta^{\mu\nu}$, the Minkowski one, so as to enable certain formal manipulations), and of the various fields or field components $\chi_a$ which enter into matter structure.  The action also depends on masses or mass-like parameters $m_a$ associated with the various $\chi_a$, as well as on essentially dimensionless coupling constants $k_b$.  Hence $S=S[g^{\mu\nu},\chi_a,m_a,k_b]$.  It will be convenient in this paper to consider each $m_a$ as a function over spacetime.

\subsubsection{Basic formula}
\label{sec:basic}  

Suppose we perform an infinitesimal continuous transformation affecting $g^{\mu\nu}, \chi_a$ and $m_a$ (but not the $k_b$).  Evidently the change in the action is
\be
\delta S = \int \left(\frac{\delta S}{\delta g^{\mu\nu}}\,\delta g^{\mu\nu} + \frac{\delta S}{\delta \chi_a}\,\delta \chi_a+\frac{\delta S}{\delta m_a}\delta m_a  \right)d^4x\,.
\label{dS}
\ee
As is well known, the equations of motion of the $\chi_a$ are given by the vanishing functional derivatives
\be
\delta S/\delta \chi_a=0.
\label{eqmotn}
\ee
If the transformation is performed on-shell (on a solution of the equations of motion of the $\chi_a$), \Eq{eqmotn} lets us drop the second term in the integral of \Eq{dS}.   It is also well known that the energy momentum tensor of the fields $\chi_a$ is given in general by
\be
T_{\mu\nu}=-\frac{2}{(-g)^{1/2}}\frac{\delta S}{\delta g^{\mu\nu}}\ .
\label{T}
\ee
(We write $T$ rather than $\Theta$ since our discussion here goes beyond ordinary matter, including, for example, radiation fields).

Now restrict the transformation we have in mind to an extended conformal transformation (ECT).  An ECT transforms $g^{\mu\nu}$ into $\Omega^{-2}g^{\mu\nu}$ (and consequently $g_{\mu\nu}$ into $\Omega^{2}g_{\mu\nu}$) for some positive spacetime function $\Omega$.  It also changes $\chi_a\to \chi_a\,\Omega^\zeta$ with $\zeta$ a real fraction depending on the type of field, and it takes each mass $m_a$ into $\Omega^{-1}m_a$ (this being allowed since, as stipulates earlier, we regard the masses as spacetime functions).  The motivation for these rules comes from dimensionality.  If the metric, rather than the coordinates, is regarded as the carrier of the length dimension $L$, then $g^{\mu\nu}$ is of dimension $L^{-2}$ while masses are of dimension $L^{-1}$ (the $\hbar$ as usual is not involved in such considerations; it is essentially dimensionless).  For infinitesimal ECTs, $\Omega=1+\delta\Omega$ with $|\delta\Omega|\ll 1$.

Thus whenever the action turns out to be invariant under an ECT, it follows from \Eqs{dS}- (\ref{T})  that
\be
0=\int \left(T_{\mu\nu}\,g^{\mu\nu}(-g)^{1/2}-m_a \frac{\delta S}{\delta m_a}\right)\delta\Omega\, d^4x.
\ee
Since $\delta\Omega$ is an arbitrary function this implies
\be
T\equiv T_{\mu\nu}\,g^{\mu\nu}=m_a\, \frac{\delta S}{\delta m_a}(-g)^{-1/2}=m_a\, \frac{\partial \mathcal{L}}{\partial m_a}\,,
\label{basic}
\ee
where $\mathcal{L}$ is the Lagrangian density: $S=\int\mathcal{L}\,(-g)^{1/2}d^4x$.   This is the key result here.  For example, it is immediately clear from it that classical massless fields with ECT invariant actions do not contribute to $T$.

\subsubsection{ECT invariance of gauge field action}
\label{sec:gauge}

An important constituent of matter is the electromagnetic field.  It derives from a vector potential $A_\mu$ which, when subjected to gauge transformations $A_\mu\to A_\mu+ \partial_\mu \Lambda$,  leaves the Faraday or electromagnetic field tensor $F_{\mu\nu}=\partial_\mu A_\nu-\partial_\nu A_\mu$ invariant.  If we agree that under an ETC $A_\mu\to A_\mu$, then the generally covariant Maxwell action,
\be
S_{M}=-\frac{1}{16\pi}\int g^{\mu\nu}\, g^{\alpha\beta}\,F_{\mu\alpha}\,F_{\nu\beta}\,(-g)^{1/2} d^4x\,,
\ee 
is obviously ETC invariant since $(-g)^{1/2}\to (-g)^{1/2}\,\Omega^4$.  We thus easily recover the widely known fact that the classical electromagnetic field does not contribute to the trace of $T_\mu{}^\nu$.

The non-Abelian gauge fields responsible for the weak and strong interactions have actions which are sums of terms like $S_{M}$, but with each term coming from a different potential, $A_\mu^\mathrm{a}$ (here ``$\mathrm{a}$'' is the symmetry group index; it labels the various potentials).  The field tensors are defined by $F^\mathrm{a}_{\mu\nu}\equiv \partial_\mu A^\mathrm{a}_\nu-\partial_\nu A^\mathrm{a}_\mu + g\,f^\mathrm{abc}A_\mu^\mathrm{b} A_\mu^\mathrm{c}$, where $g$ is the dimensionless gauge coupling constant and $f^{abc}$ (antisymmetric in $\mathrm{b}$ and $\mathrm{c}$)  is the collection of dimensionless structure constants characterizing the gauge symmetry group in question.    As with the Maxwell case we take it that under ECTs $A^\mathrm{a}_\mu\to A^\mathrm{a}_\mu$.  But we require that $F^\mathrm{a}_{\mu\nu}\to F^\mathrm{a}_{\mu\nu}\,$ in order to obtain ECT invariance.  It is thus consistent to assume that under an ECT $g\to g$ and $f^\mathrm{abc}\to f^\mathrm{abc}$, as would also be suggested by their dimensionality.  Having verified the ECT invariance of the gauge field's action, we see that classically non-Abelian gauge fields do not contribute to the total $T_\mu{}^\mu$. 

\subsubsection{ECT invariance of fermion field action}
\label{sec:fermion}

Ordinary matter is made of fermions; these are Dirac particles, e.g. electrons, or bound states of Dirac particles, e.g. nucleons.  We now delve into the machinery required to demonstrate the ECT invariance of the Dirac field action~\cite{Bek80} which will have consequences for the central question we study here. For reasons already mentioned, the physics has to be formulated in curved space.  The Dirac field is a 4-spinor field $\Psi$.  One defines~\cite{Wheeler,DeWitt} four $4\times 4$ matrix fields $\gamma^\mu$ by $\gamma^\mu=\lambda^\mu_A \,\gamma^A$ where the $\gamma^A; A=0,1,2,3$ are the ordinary (constant and dimensionless) Dirac matrices obeying $\gamma^A\gamma^B+\gamma^B\gamma^A=2\eta^{AB}I$ with $I$ the $4\times 4$ unit matrix.   The $\lambda^\mu_A$ are the four 4-vectors forming an orthonormal tetrad that underlies the metric: $g^{\mu\nu}=\eta^{AB}\,\lambda^\mu_A\,\lambda^\nu_B$.  Accordingly $\gamma^u\gamma^\nu+\gamma^\nu\gamma^\mu=2g^{\mu\nu}I$ with $I$ the $4\times 4$ unit matrix. These relations make it clear that under ECTs $\lambda^\mu_A\to\lambda^\mu_A\,\Omega^{-1}$ and $\gamma^\mu\to\gamma^\mu\,\Omega^{-1}$. (By convention we take the Minkowski $\eta^{AB}=\mathrm{diag}(-1,1,1,1)$ as invariant under ECTs). 

The spinor-affine connection $\Gamma_\mu$ is defined by 
\be
\Gamma_\mu\equiv \fourth \gamma^A \gamma_\nu \lambda^\nu_A{}_{;\mu}\,,
\label{Gamma}
\ee
As would be expected of any connection, under ECTs $\Gamma_\mu$ transforms inhomogeneously, i.e.,
 \be
 \Gamma_\mu\to\Gamma_\mu+\fourth(\gamma^\nu\gamma_\mu-\gamma_\mu\gamma^\nu)\Omega^{-1}\partial_\nu\Omega\,.
 \label{Gammato}
 \ee
In curved spacetime the familiar Dirac matrix $\beta$ is replaced by a $4\times 4$ matrix field $\gamma$ satisfying $\gamma\gamma_\mu+\gamma_\mu{}^\dagger\gamma=0$ as well as $\partial_\mu\gamma+\gamma\Gamma_\mu+\Gamma_\mu{}^\dagger\gamma=0$.   We assume that $\gamma^\dagger=\gamma$; this does not contradict either of these equations (and agrees with the customary flat spacetime choices for $\beta$).   It is consistent with these equations and dimensionality to assume that $\gamma\to \gamma$ under ECTs. 

The generally covariant action of the Dirac field is written as~\cite{Wheeler,DeWitt} 
\be
S_D=\int\Psi^\dagger\gamma\big[\imath\hbar\gamma^\mu(\partial_\mu\Psi-\Gamma_\mu\Psi)-m_f\Psi\big](-g)^{1/2}d^4x\,,
\label{SD}
\ee
 where the ordinary derivative of the flat spacetime equation is transformed into a covariant derivative with help of $\Gamma_\mu$, and $m_f$ is the fermion's mass.  It is clear that the derivative term in the Lagrangian will be invariant under a special ETC with \emph{constant} $\Omega$ only if we assume that $\Psi\to\Psi\Omega^{-3/2}$.  This choice will also guarantee that the mass term is invariant (recall we stipulated $m_f\to m_f\,\Omega^{-1}$).  Since $\Gamma_\mu$ would not change under such special ECT we obtain invariance of the action.  So let us adopt the rule $\Psi\to\Psi\Omega^{-3/2}$ also for spacetime dependent $\Omega$.  In the latter case we need to take into account the gradient of $\Omega$ arising from $\partial_\mu\Psi$ as well as from the change in $\Gamma_\mu$ following an ECT.  A tedious but straightforward calculation using the anticommutation rule for the $\gamma^\nu$ shows that the $\partial_\mu\Omega$ terms all cancel.  Thus the massive Dirac field classical action  is invariant under ECTs~\cite{Bek80}.  
 
 This statement remains correct when the Dirac field is coupled to gauge fields.  In that case we would replace in the action $\partial_\mu\Psi\to \partial_\mu\Psi-\imath (e/\hbar) A_\mu\Psi$  for the electromagnetic field or $\partial_\mu\Psi\to \partial_\mu\Psi-\imath (g/\hbar)  T^a A^a_\mu\Psi$ for a non-Abelian gauge field, e.g. gluon field.  Here $g$ is the gauge coupling constant already introduced above, and the $T^a$ are the \emph{dimensionless} matrices of the particular representation of the gauge symmetry group being employed.  Because $e, g$ and $T^a$ can be assumed invariant under ECTs, the above argument continues to work.
 
In quantum chromodynamics, the theory behind the existence of hadrons as quark composites, the quark masses $m_f$ in \Eq{SD} are actually induced by the vacuum expectation value of the (scalar and possibly complex) Higgs field $\Phi$ after spontaneous symmetry breaking (SSB).  Likewise for the mass of the electron.  So we must first demonstrate that Higgs field physics is ECT invariant and gives rise to fermion masses that may be considered as transforming as $m_f\to m_f\,\Omega^{-1}$.

\subsubsection{ECT invariance of Higgs field physics}
\label{sec:Higgs}
 
 The standard action for the (scalar) Higgs field is~\cite{Ryder}
 \be
 S_H=\int \Big[-\half g^{\mu\nu}\partial_\mu\Phi\,\partial_\nu\Phi+\half m_H^2\hbar^{-2} \Phi^2-\fourth\lambda\Phi^4 \Big](-g)^{1/2}d^4x
 \label{SH} 
 \ee
where $\lambda$ is the dimensionless self-coupling constant and $m_H$ is the Higgs mass scale (not the mass of the field as the term has the ``wrong'' sign).  In reality we should have written $\Phi^\dagger$ along $\Phi$ because the Higgs field is generally invoked in a nontrivial representation of the relevant symmetry group, e.g. SU(3) in quantum chromodynamics, and so is really a column ``vector'' containing several possibly complex components.  Since this complication is inessential for the issues before us, we pretend the Higgs field is in the trivial representation, and write $\Phi$ as a single real field.  Before SSB the factor $m_f\Psi^\dagger\gamma\Psi$ in \Eq{SD} would actually be the Yukawa type term $k \Psi^\dagger\gamma\Phi\Psi$ with $k$ a suitable dimensionless coupling constant.

Because $\lambda$ and $k$ are dimensionless, we prescribe $\lambda\to\lambda$ and $k\to k$ under ECT.  In harmony with $m_H$'s dimension $[\hbar/\mathrm{length}]$ we must prescribe $m_H\to m_H\,\Omega^{-1}$.  Thus in order not to spoil the ECT invariance of $S_D$ with the Higgs field included, it is necessary to assume that it transforms as $\Phi\to\Phi\,\Omega^{-1}$.  Returning to \Eq{SH} we see that $S_H$ is invariant under a ECT with constant $\Omega$, but when $\Omega$ depends on the coordinates, the kinetic part of the Lagrangian density produces terms containing $\partial_\mu\Omega$  that spoil the invariance.  An ECT invariant action---also for variable $\Omega$---is had by including in the integrand of $S_H$ the term $-\frac{\scriptstyle 1}{\scriptstyle 12}R\Phi^2$, with $R$ in this section only being the Ricci scalar deriving from the metric~\cite{Penrose}.  Of course this addition affects the usual scalar energy-momentum tensor, and in fact lends it improved properties~\cite{Jackiw}.

SSB forces $\Phi$ to zero into its vacuum expectation value at which the ``potential'' in $S_H$ reaches its minimum.  Ignoring the spacetime curvature correction this value is $(m_H/\hbar\surd\lambda)$; it replaces $\Phi$ in the term $k \Psi^\dagger\gamma\Phi\Psi$ that originally appears in the Dirac action.  Thus the fermion in question acquires a mass $m_f=k\, (m_H/\hbar\surd\lambda)$ which evidently scales as $\Omega^{-1}$ under an ECT.  All our remarks about $S_D$'s invariance thus remain valid.

The difference $\eta\equiv \Phi-(m_H/\hbar\surd\lambda)$ represents the dynamic residue of the Higgs field.  Substituting $\Phi=(m_H/\hbar\surd\lambda)+\eta$ into the integrand of $S_H$ yields, first of all, the constant term $\fourth (m_H^4/\hbar^4\lambda)$ which can be interpreted as a correction to the cosmological constant, and the term $-\frac{\scriptstyle 1}{\scriptstyle 12}(m_H^2/\hbar^2\lambda) R$ which contributes a correction to the gravitational constant in general relativity.  We ignore these two terms since they are germane only to the gravitational physics.  The action for $\eta$ thus works out to be to
\be
S_\eta=\int \Big[-\half g^{\mu\nu}\partial_\mu\eta\,\partial_\nu\eta-\frac{\scriptstyle 1}{\scriptstyle 12}R\,\eta^2-(m_H^2/\hbar^2\lambda)\eta^2+\mathcal{O}(\eta^3) \Big](-g)^{1/2}d^4x\,.
\label{eta}
\ee
Corrections of $\mathcal{O}(\eta^3)$ excluded,  $\eta$'s action is that of a conformally coupled scalar field with mass term of the customary sign (the mass is $\surd 2\, m_H/\surd\lambda)$.  Before quantum corrections it retains the ECT invariance of its precursor, the curvature corrected $S_H$.

Gluon gauge fields do not couple to an Higgs field; they thus remain massless after SSB.
In principle we should also have investigated the interaction between Higgs field and those gauge fields which transmute via SSB into the intermediate massive vector bosons of the Weinberg-Salam electro-weak theory.  However, since we are only interested in ordinary matter made of atoms (electrons and nucleons) in whose structure weak interactions play an insignificant part (energetically speaking), we shall forego that analysis which entails some tricky issues. 

\subsubsection{Argument for the subdominant trace energy condition for ordinary matter}
\label{sec:proof}

The energy momentum tensor of ordinary matter, $\Theta_{\mu\nu}$, comes from the actions for electromagnetism, non-Abelian gauge (gluon) fields, Dirac (electron and quark) fields, and the Higgs field.  We showed that at the classical level all these ingredient enjoy symmetry under ECTs, before and after SSB.  The mentioned actions are functionals of the metric and dynamic fields, and functions of the various masses.   They depend as well as on the constants $e, g, k, \lambda$ and $f^{abc}$ which, as explained earlier at various points, are ECT invariant.  Hence, when we consider the variation of the action under ECT we can ignore them, but not the masses.  This is the justification for using \Eq{basic} for $\Theta_{\mu\nu}$.   At the classical level we learn from \Eq{basic} that the trace $\Theta=\Theta_\mu{}^\mu$ only gets contributions from the mass terms of actions.  

All Dirac actions are linear in the masses; hence their contributions to $\Theta$ are just the mass dependent terms in the action integrand, namely $-m_f\Psi^\dagger\gamma\Psi$ for each fermion.  Since we are primarily interested in results in flat spacetime we can take $\gamma$ as the usual Dirac matrix $\beta={\rm diag}(1,1,-1,-1)$. Let us break up $\Psi$ into an upper 2-spinor $\psi$ and a lower one $\phi$. Thus $\Psi^\dagger\gamma\Psi=\psi^\dagger\psi-\phi^\dagger\phi$.  As we know from quantum mechanics, the $\phi$ represents negative energy (or antiparticle) components, and for a nonrelativistic fermion it is small compared to $\psi$, so that $\Psi^\dagger\gamma\Psi$ is evidently positive.  This tells us that atomic and free electrons contribute \emph{negatively} to $\Theta$. But for a nucleon, in which the u and b quarks are evidently relativistic, the above argument is less credible.  Yet the fact remains that in ordinary matter nucleons are nonrelativistic.  Treated as a Dirac particle each would thus contribute positively to $\Psi^\dagger\gamma\Psi $.  This is a good argument for the negativity of the nucleon contributions (of form $-m_f\Psi^\dagger\gamma\Psi$ after SSB) to $\Theta$.

What about the Higgs field itself?  After SSB the relevant action is $S_\eta$.  Since the mass term is quadratic in $m_H$, the only mass parameter, the contribution to $\Theta$ must be twice the mass term, $-(m_H^2/\hbar^2\lambda)\eta^2$, which is obviously negative.  We  have already mentioned that gluon fields do not contribute to $\Theta$.  Hence in the classical description of ordinary matter the $\Theta$ must be negative.

Field quantization introduces complications to the above conclusion in the form of the Weyl (or conformal or trace) anomaly.  For massless fields the conformal invariance of the classical theory is no longer a symmetry of the quantum field theory~\cite{Davies}.  As a result the \emph{expectation value} of the trace of the energy momentum tensor of a massless conformal field, e.g. the electromagnetic field, receives a  contribution (usually independent of the quantum state) consisting of a polynomial in curvature invariants (with no zero order term) together with a second derivative of $R$~\cite{Davies}.   Since we ultimately  have in mind a flat spacetime problem, we can ignore the anomalous curvature terms and the expectation value of $\Theta$ continues to vanish.   For \emph{massive} scalar (with or without curvature coupling) and Dirac fields, the expectation value of $\Theta$ contains~\cite{Parker}, apart from a polynomial in curvature and second derivatives of $R$, a term proportional to $-m^4$, where $m$ is the mass of the particular field (the sign here has been adjusted for the difference between Parker and Toms' signature convention and ours). Thus, whether classically or quantum mechanically, ordinary matter in (nearly) flat spacetime has a negative $\Theta$.

It must be emphasized that the above arguments do not yet constitute a proof.  Ordinary matter is a bound state of Dirac, Higgs and gauge fields.  We have just argued that each of these fields individually contributes negatively to $\Theta$.  But it may be that the effective theory of ordinary matter that supplants this field-by-field description is sufficiently different from the ingredient theories so as not to respect the $\Theta<0$ rule.  Or put more colorfully, the whole may be more than the sum of the parts.  Nevertheless, in view of the convergence of conclusions from the kinetic and field-theoretic arguments, it seems safe to rely on the conclusion that $\Theta<0$ for ordinary matter.

This means that $-\Theta_{tt}+\Theta_i{}^i<0$.  Since we assume positive energy density for ordinary matter, $\Theta_i{}^i$ must either be negative, or if positive it must be bounded by $\Theta_{tt}$.   We have already inferred in Sec.~\ref{sec:Def} from the strong energy condition, \Eq{cond2},  that when $\Theta_i{}^i<0$ we have $|\Theta_i{}^i|<T_{tt}$.  Accordingly, we have just confirmed the subdominant trace energy condition, \Eq{cond}, for ordinary matter.  In what follows we only use the $\Theta_i^i>0$ case of the subdominant trace condition; this is independent of the strong energy condition.

\section{Demonstration of positivity of total mass}
\label{sec:two_proofs}
 
\subsection{The virtual work method}
\label{sec:way1}

Henceforth we display $c$ explicitly.
Consider first a massless quantum field in a stationary state confined to a spherical space of radius $R$  by virtue of some boundary conditions on the enclosing wall.  (We no longer mention the scalar curvature, so there is no risk of confusion with its symbol $R$.) We assume that these conditions do not introduce an extra scale of length.  Examples would be Dirichlet or Neumann conditions for a scalar field and perfectly conducting walls for the electromagnetic field.  On dimensional grounds the field mode frequencies in the space must take the form
$ \omega_j=\epsilon_j c/R $, with the constants $\epsilon_j$ dimensionless and positive.  Therefore,
the one-particle energy eigenvalues are\begin{equation}
E_j=\epsilon_j\hbar c/R\,; \qquad j=1,2,\cdots
\label{spect}
\end{equation}
Likewise, the vacuum energy must be of the form
\begin{equation}
E_{\rm v}=\alpha\hbar c/R\,,
\label{Evac}
\end{equation}
with $\alpha$ a dimensionless real constant.   These forms are verified in all known examples~\cite{Boyer,Wolfram,Mostep}.  

If the wall is devoid of symmetry, or has lower than spherical symmetry,  \Eq{Evac} must be replaced by
\be
E_{\rm v}=\frac{\alpha(\xi_1,\xi_2,\cdots)\hbar c}{R}\,.
\label{Evac2}
\ee
Here $R$ is a relevant length, e.g. the radius of a sphere just circumscribing the space in question.  The dependence $E_{\rm v}\propto R$ follows, again, on dimensional grounds since the only dimensional natural constants of relevance are $c$ and $\hbar$.  But now we must include a number of dimensionless parameters, $\xi_1,\xi_2,\cdots$, which specify the shape of the boundary of the space which defines the vacuum state, e.g. axes ratios for an ellipsoidal space.  There could be many such $\xi$'s, particularly when there is no symmetry at all, or when the surface has features on many diverse length scales.

We now imagine a virtual homologous dilation of the system: $R\to R+\delta R$ (with the $\xi$'s fixed).  Obviously the change in vacuum energy is
\be
\delta E_{\rm v}=-E_{\rm v}\,\frac{\delta R}{R}\,.
\label{delE}
\ee
At the same time the energy of the wall changes due to its stretching.  To calculate this last energy change we proceed as follows. 

Assuming the wall is not necessarily infinitesimally thin, we parcel it into small 3-dimensional cells.  If these are sufficiently small, we can make them cubic in shape.  Of course this cannot be done consistently on a global scale: the curvature of the wall will require, in addition to cubic cells, cells of different shapes to fill in the gaps between them.  But it should be obvious that the volume taken up by the latter cells can be made relatively negligible as the cubic cells become smaller.  We suppose the typical edge of each cube is of length $a$. The local orientation of the cubic cells (in any small region) is to be determined as follows.  

The stresses of the wall structure are described by the space-space part of its energy-momentum tensor $\Theta_{\mu\nu}$, namely $\Theta_{ij}$. Because  $\Theta_{ij}$ is symmetric we can diagonalize it separately at each point of the wall.  It will have three real eigenvalues, $\tau_1,\tau_2$ and $\tau_3$, and three associated  \emph{orthonormal} eigenvectors, $\vec v_1, \vec v_2$ and $\vec v_3$, respectively.  The cubes we mentioned will be oriented so that their faces are orthogonal to the local $\vec v_1, \vec v_2$ and $\vec v_3$.

Recall that $\Theta_{ij}\,n_j$ is the flux of the $i$ component of momentum in the direction of the unit vector $\vec n$.    Thus the vector forces acting \emph{on} the cube's six faces, taken in appropriate order,  are $a^2\tau_1 \vec v_1,\ a^2\tau_2 \vec v_2,\ a^2\tau_3 \vec v_3,\ -a^2\tau_1 \vec v_1,\ -a^2\tau_2 \vec v_2$ and $-a^2\tau_3 \vec v_3$ (we ignore the variation of the various quantities across the cube).  The sign here attached to $\vec v_1, \vec v_2$ or $\vec v_3$, etc. is determined by the requirement that that vector point \emph{into} the cube.  Upon the alluded dilation  $a\to a+\delta a$ and  the areas of the cube's faces will be stretched to $a^2+2a\,\delta a$ while the eigenvalues and eigenvectors will change to $\tau_i(1+\mathcal{O}(\delta a))$ and $\vec v_i(1+\mathcal{O}(\delta a))$, respectively.  Evidently the virtual work \emph{done on} the cube by the dilation is
\be
\delta w = -\half\delta a (a^2\tau_1+a^2\tau_2+a^2\tau_3+a^2\tau_1+a^2\tau_2+a^2\tau_3)+\mathcal{O}(a\,\delta a^2)=-(\tau_1+\tau_2+\tau_3)a^2\,\delta a +\mathcal{O}(a\,\delta a^2)
\ee 
with the factor $\half$ coming from the fact that each face of the cube is displaced parallel to itself by $\half \delta a$, and the negative sign reflecting the fact that the cube \emph{expands} while the forces on it are reckoned as acting on it.  Of course $\delta a/a=\delta R/R$.
Thus to dominant order
\be
\delta w=-(\tau_1+\tau_2+\tau_3)\,a^3\,\delta R/R = -\Theta_i{}^i\,a^3\,\delta R/R\,,
\ee
where a sum over $i$ is understood.

Finally, adding the contributions from all the cubes we have for the  {\em total} virtual work $W$  performed {\em on} the surface structure the integral over the wall's volume
\be
\delta W  =- \frac{\delta R}{R}\int_{\rm w} \Theta_i{}^i\, d^3x\,.
\ee
Of course $\Theta_i{}^i$ may vary from point to point of the wall.

In an equilibrium situation the variation of the {\em total} energy under any virtual change of size must vanish: $\delta E_{\rm v} + \delta W=0$.  In particular, this variation must vanish under the virtual dilation of the system studied above.  In view of \Eq{delE} the condition for equilibrium is
\be
 E_{\rm v}
=-\int_{\rm w} \Theta_i{}^i\, d^3x\,.
\label{Ev}
\ee
But the subdominant trace condition (just for $\Theta_i^i>0$) tells us that
\be
-\int_{\rm w} \Theta_i{}^i\, d^3x>-\int_{\rm w} \, \Theta_{tt}\, d^3x=-M_\mathrm w
\label{final}
\ee
Hence the total mass-energy of the system, $E_{\rm v}+M_{\rm w}$, is positive, even when $E_{\rm v}$ is negative.

\subsection{The trace method}
\label{sec:way2}

In the previous section we characterized the confined quantum field as devoid of a scale (such as mass).  Here we shall characterize it as just possessing symmetry under ECTs.  The relevant class of fields obviously includes the Maxwell field and the gluon non-Abelian gauge field $A^a_{\mu}$  (see Sec.~\ref{sec:gauge}).    A second example would be a \emph{massive} scalar field or field multiplet $\Phi$ conformally coupled to curvature as illustrated by the quadratic part of the action (\ref{eta}), and optionally self-coupled through the $\lambda \Phi^4$ term of \Eq{SH}.   Yet a third example is furnished by the Dirac field $\Psi$ surveyed in Sec.~\ref{sec:fermion}.   

Of course if $m_f=0$ this last is really a chiral neutrino field, so only its left handed part,  $\Psi_l$, is physical.  In curved spacetime (required to make sense of conformal transformations) one would separate out the left part as follows:
\be
\Psi_l=(I-\gamma^5)\Psi;\quad  \gamma^5=c(\imath/4!)\epsilon_{\mu\nu\rho\sigma}\gamma^\mu\gamma^\nu\gamma^\rho\gamma^\sigma\,.
\ee  
Recall that the Levi-Civita tensor $\epsilon_{\mu\nu\rho\sigma}$ carries a factor $(-g)^{1/2}$ which scales as $\Omega^{4}$.  Concomitantly, each $\gamma^\mu$ scales with a factor $\Omega^{-1}$ (see Sec.~\ref{sec:fermion}).  Hence both $\gamma^5$ and the left-part projecting operator are unchanged under the ECT transformation.  Thus the chiral (or Weyl) neutrino field theory inherits the ECT invariance of the generic Dirac theory.

In this section the quantum field's energy-momentum tensor is denoted by plain $T_\mu{}^\nu$.  Returning to \Eq{basic} it is clear that classically $T\equiv T_\mu{}^\mu$ gets no contribution from the electromagnetic, gauge or neutrino fields. But as discussed in Sec.~\ref{sec:proof}, classical massive scalar and Dirac fields will contribute negatively.  (One might raise a quibble concerning the massive Dirac field, the sign of $\Psi^\dagger\gamma\Psi$ not being certain for a relativistic field.  However, with the appearance of the conformal anomaly the consequences for the sign of $T$ are clear, as detailed in Sec.~\ref{sec:proof}.)  

Two points must be emphasized.  First, the mentioned result refers to the  field energy-momentum tensor of all fields that interact with one another, not necessarily to one field's energy-momentum tensor, unless that particular field does not interact with others (apart from gravitationally).  Second, as already mentioned in Sec.~\ref{sec:proof}, the conformal anomaly puts in an appearance, so that the expectation value $\langle T_\mu{}^\mu\rangle$ depends on the local curvature invariants and, for massive spinor and scalar fields, receives an extra negative contribution depending only on the field's mass.  Since we ultimately work in flat spacetime, we can discard the curvature terms.  For simplicity in what follows we drop the $\langle\cdots\rangle$ notation: $T_\mu{}^\nu$ and $T$ in this section represent the \emph{expectation values} of the energy-momentum tensor and its trace, respectively.  Thus for all the quantum fields under consideration
\be
T\leq 0.
\label{T0}
\ee

Now let ${}^{\rm (tot)}T{}^\mu{}_\nu$ stand for the total energy-momentum tensor including wall and field contributions.  From energy-momentum conservation, or from Einstein's equations in the flat-spacetime limit, it follows that $\partial_\mu\,{}^{\rm (tot)}T{}^\mu{}_\nu=0$.  In a stationary situation, such as contemplated here, there is no time dependence, so a spatial component of the last equation may be written as $\partial_j\,{}^{\rm (tot)}T{}^j{}_i=0$.  Our use of ordinary partial derivatives emphasizes that we work in Cartesian coordinates.  Let us multiply the above equation by $x^i$, sum over $i$, integrate over all space and integrates by parts.  The result is
\be
\oint_\infty {}^{\rm (tot)}T^j{}_i\ x^i\ d\sigma_j = \int {}^{\rm (tot)}T{}^j{}_i\ \delta_j^i\,d^3x= \int {}^{\rm (tot)}T^i{}_i\,d^3 x.
\ee 
The boundary integral must vanish because the wall's energy momentum tensor is localized, while the field's vacuum energy momentum tensor should decay asymptotically once all usual divergences are regularized and renormalized.  We conclude that the space integral of 
${}^{\rm (tot)}T^i{}_i$ vanishes.  This immediately gives
\be
\int T^i{}_i\,d^3 x = -\int \Theta^i{}_i\,d^3 x.
\label{inter}
\ee

We now write the vacuum energy as
\be
E_{\rm v}=\int T_{tt}\,d^3x.
\label{Evac1}
\ee
The integral here extends over all space: it is well known that vacuum energy \emph{density} is not actually restricted to the cavity but receives a contribution from the region beyond the wall.  Now we can use \Eq{T0} to write
\be
\int T_{tt}\,d^3x = -\int T^t{}_t\,d^3x \geq  \int T^i{}_i\,d^3x.
\ee
Substituting \Eqs{inter}-(\ref{Evac1})  in this and employing the subdominant trace condition (again only for $\Theta_i^i>0$)  we have
\be
E_\mathrm{v}=\int T_{tt}\,d^3x \geq -\int \Theta^i{}_i\,d^3 x >  -\int \Theta_{tt}\, d^3x = -M_\mathrm {w}
\ee
which amounts to $E_{\rm v}+M_{\rm w}>0$.  Thus we have again demonstrated positivity of the total mass-energy without  having to consider the variation of energy of the vacuum or the work done on the wall during an homologous deformation. 

\section{Summary and Caveats}

We have shown why a composite cavity-quantum field system cannot have negative mass energy.  Key to our argumentation is the subdominant trace energy condition which is suggested by kinetic theory arguments and has here been buttressed by arguments from classical and quantum field theory.  The fields which are found to satisfy the condition are those endowed with ECT, an extended form of conformal symmetry which does not preclude the presence of rest masses.

This already tells us that the result obtained here---positivity of total mass---is by no means generic.  It pertains to a rather large class of quantum fields, massless or massive, confined by cavities constructed out of baryon-electron matter.  But it may not be provable, at least not by the means employed here, for a cavity made of exotic matter, e.g. a soliton of a scalar field with higher order nonlinear self-interaction.

The last caveat concerns interactions.  Minimal coupling interactions with gauge fields and the Yukawa interaction with the Higgs field have here been included in the Dirac action, so our arguments about the sign of $T$ or $\Theta$ would seem to be unaffected by such interactions (though there is a dearth of conformal anomaly calculations for gauge coupled fermions).  However, an interaction which shows up as an extra term in the action which cannot be incorporated naturally into one of the field  actions discussed above, would usually generate an extra contribution to the energy momentum tensor.  The tensor could then no longer be cleanly split into $\Theta_{\mu\nu}$ and $T_{\mu\nu}$, as required by both arguments in Sec.~\ref{sec:two_proofs}, thus compromising our conclusion. 

\section*{Acknowledgments}

I thank Matt Visser for much lore on energy conditions.  This research was supported by the I-CORE Program of the Planning and Budgeting Committee and the Israel Science Foundation (grant No. 1937/12), as well as by the Israel Science Foundation personal grant No. 24/12.

\end{document}